\documentclass[runningheads]{llncs}

\def\BibTeX{{\rm B\kern-.05em{\sc i\kern-.025em b}\kern-.08em
    T\kern-.1667em\lower.7ex\hbox{E}\kern-.125emX}}

\usepackage[T1]{fontenc}
\usepackage{graphicx}

\usepackage{amsmath}
\usepackage{hyperref}
\usepackage{multirow}
\usepackage{makecell}
\usepackage{xcolor}
\usepackage{marvosym} 
\usepackage{floatrow}
\usepackage{booktabs} %
\usepackage{rotating}
\usepackage{todonotes}
\usepackage{orcidlink}

\begin{document}
\title{To Case or Not to Case: \\ An Empirical Study in Learned Sparse Retrieval}
\titlerunning{To Case or Not to Case}
\author{Emmanouil Georgios Lionis\inst{1,2,3}\orcidlink{0009-0004-3931-9657} \and
Jia-Huei Ju\inst{2}\orcidlink{0000-0003-2247-3370} \and
Angelos Nalmpantis\inst{3}\orcidlink{0000-0002-1505-4656} \and
Casper Thuis\inst{3}\orcidlink{0009-0006-8602-9665} \and
Sean MacAvaney\inst{1}\orcidlink{0000-0002-8914-2659} \and
Andrew Yates\inst{4}\orcidlink{0000-0002-5970-880X}
}
\authorrunning{E. Lionis et al.}
\institute{ 
            University of Glasgow, United Kingdom \and
            University of Amsterdam, The Netherlands \and
            TKH AI, The Netherlands \and
            Johns Hopkins University, United States \\
            \email{e.lionis.1@research.gla.ac.uk}
}

\maketitle              %
\begin{abstract}
Learned Sparse Retrieval (LSR) methods construct sparse lexical representations of queries and documents that can be efficiently searched using inverted indexes. Existing LSR approaches have relied almost exclusively on uncased backbone models, whose vocabularies exclude case-sensitive distinctions, thereby reducing vocabulary mismatch. However, the most recent state-of-the-art language models are only available in cased versions. Despite this shift, the impact of backbone model casing on LSR has not been studied, potentially posing a risk to the viability of the method going forward. To fill this gap, we systematically evaluate paired cased and uncased versions of the same backbone models across multiple datasets to assess their suitability for LSR. Our findings show that LSR models with cased backbone models by default perform substantially worse than their uncased counterparts; however, this gap can be eliminated by pre-processing the text to lowercase. Moreover, our token-level analysis reveals that, under lowercasing, cased models almost entirely suppress cased vocabulary items and behave effectively as uncased models, explaining their restored performance. This result broadens the applicability of recent cased models to the LSR setting and facilitates the integration of stronger backbone architectures into sparse retrieval. The complete code and implementation for this project are available at: \url{https://github.com/lionisakis/Uncased-vs-cased-models-in-LSR}

\end{abstract}

\section{Introduction}

Ad hoc retrieval, a central task in Information Retrieval (IR)~\cite{crestani_is_2001}, aims to return the most relevant documents in response to user queries. Traditional IR methods, such as TF-IDF~\cite{sparck_jones_statistical_1972} and BM25~\cite{robertson_probabilistic_2009}, rely on statistical feature extraction and exact lexical matching between query terms and document content. These approaches form the backbone of efficient retrieval systems, like PISA~\cite{mallia_pisa_2019}, offering both scalability and strong performance. However, because they depend on exact term matching, inconsistencies can arise when terms differ due to surface-level characteristics, such as inflection and capitalization. Most IR pipelines apply pre-processing steps such as stemming and lowercasing to mismatches between query and document terms~\cite{al-hawari_machine_2021,kunneman_question_2019}.

Recent advances in transformer-based pre-trained language models have transformed natural language understanding~\cite{devlin_bert_2019}, enabling retrieval systems to move beyond lexical overlap toward richer semantic matching. This shift has spurred the development of Neural Information Retrieval (Neural IR)~\cite{guo_deep_2016,hambarde_information_2023}. Within this paradigm, \textbf{Learned Sparse Retrieval} (LSR)~\cite{formal_splade_2021,nguyen_adapting_2023,zamani_neural_2018,zhuang_tilde_2021} has emerged as a particularly effective approach. LSR leverages the semantic knowledge of pre-trained models to assign learned weights to lexical terms, encoding queries and documents as sparse vectors over a fixed vocabulary. While most vector dimensions have a weight of zero, the non-zero values capture term importance, analogous to concepts such as term frequency or TF-IDF. This representation preserves compatibility with inverted index infrastructures~\cite{mitra_introduction_2018,lin_proposed_2021,mallia_learning_2021} while offering greater transparency by explicitly highlighting the terms that contribute to relevance scoring.

Built on top of pre-trained language models, the effectiveness of LSR methods is closely tied to the specific model variant employed. Most early pre-trained models were released in both \textit{cased} and \textit{uncased} forms~\cite{devlin_bert_2019,sanh_distilbert_2020}. Cased models preserve capitalization, distinguishing tokens such as \textit{Apple} and \textit{apple}. This is advantageous in tasks, such as Named Entity Recognition (NER) ~\cite{lee_biobert_2020}, where case is an important signal, often influencing semantic meaning. In contrast, uncased models convert all text to lowercase, merging both forms into a single token. This approach reduces data sparsity and aligns with the lowercasing practices common in traditional IR pipelines~\cite{al-hawari_machine_2021,kunneman_question_2019}, making it effective for contexts where case distinctions are less critical (e.g., open-domain question answering~\cite{karpukhin_dense_2020}). However, removing casing also discards potentially meaningful differences, such as those between the company \textit{Apple} and the fruit \textit{apple}.

Despite these trade-offs, prior work in LSR~\cite{formal_splade_2021,nguyen_adapting_2023} has predominantly relied on uncased models, with limited systematic investigation into how casing choices affect retrieval effectiveness. In contrast, several recent studies~\cite{qiao_leveraging_2025,xu_csplade_2025,zeng_scaling_2025,doshi_mistral-splade_2024} have explored cased models, specifically decoders such as T5~\cite{raffel_exploring_2023} and OPT~\cite{zhang_opt_2022}, without considering their uncased counterparts. This leaves an important gap in understanding the behavior of cased versus uncased models and complicates the use of state-of-the-art architectures, such as ModernBERT~\cite{warner_smarter_2024}, which are released only in cased forms. These practical challenges underscore the need for a careful examination of casing in LSR. Given the strong influence of model vocabulary on the output representations of LSR models, coupled with the language modeling community's increasing shift toward cased-only models, it is important to systematically investigate the role of casing to ensure the long-term viability of LSR methods.

In this work, we aim to address this gap by investigating the following research questions (RQs):

\begin{enumerate}
\label{RQs}
    \item[] \textbf{RQ1:} \label{RQ1} How does the casing of backbone models impact the performance of LSR?
    \item[] \textbf{RQ2:} \label{RQ2} Can lowercasing pre-processing improve the performance of cased models?
    \item[] \textbf{RQ3:} \label{RQ3} Can post-processing strategies, such as constraining logits or modifying training to reduce capitalization reliance, improve cased models?
    \item[]\textbf{RQ4:} \label{RQ4} What is the effect of backbone model casing on zero-shot task transfer?
\end{enumerate}

Our experiments demonstrate that the choice of cased versus uncased models has a substantial impact on LSR performance. Specifically, uncased models generally outperform cased models when no pre-processing is applied. However, applying lowercasing pre-processing to cased models effectively mitigates this gap, enabling them to achieve performance comparable to uncased models. Post-processing techniques, such as using only the uncased tokens or regularizing cased tokens, provide modest efficiency gains while causing only a slight decrease in accuracy. Additionally, we analyze the distribution of token types (cased vs. uncased) in model outputs based on the input token type. The results show that, without pre-processing, cased models frequently map cased inputs to uncased tokens, indicating a systematic preference for the uncased vocabulary. With lowercasing, cased models rely almost entirely on uncased tokens, effectively mirroring the behavior of uncased models. These findings demonstrate that while casing introduces challenges for LSR, appropriate pre-processing enables cased models to remain competitive and effective for both in-domain and out-of-domain retrieval tasks.

\section{Related Work}\label{sec:related}

\subsection{Learned Sparse Retrieval}
LSR methods encode queries and documents into sparse representations with two encoders, a query encoder ($f_Q$) and a document encoder ($f_D$). These two encoders, which can be the same or distinct, encode text as a high-dimensional sparse vector with a dimension $|V|$ (matching the size of the model's vocabulary, typically around $30$k) where the majority of the elements are zero.\footnote{To induce this sparsity in the output representations, there have been numerous pruning and regularization methods, like top-k pruning~\cite{macavaney_expansion_2020}, FLOPS regularization~\cite{paria_minimizing_2020}, and DF-FLOPS regularization~\cite{DBLP:conf/sigir/PorcoMMRKPM025}.} Document relevance is computed using the dot product between the query and document vectors:
\begin{equation}
score(q, d) = f_Q(q) \cdot f_D(d) = w_q \cdot w_d = \sum^{|V|}_{i=1} w_q^i w_d^i,
\label{eq:LSR-score}
\end{equation}
where $w^i$ denotes the predicted weight for the $i^{th}$ term in the vocabulary.

The primary distinction between LSR and traditional methods like BM25 lies in their approach to tokenization and term weighting. In terms of tokenization, traditional lexical models use a rule-based pre-processing pipeline (usually including steps like stopword removal, stemming, and lowercasing) while LSR usually uses the backbone language model's tokenizer (usually a WordPiece~\cite{DBLP:conf/icassp/SchusterN12} or BPE~\cite{DBLP:conf/acl/SennrichHB16a} model). For term weighting, BM25 relies on corpus-level statistics (e.g., TF-IDF) and does not require training, while LSR learns term weights through semantic processing and supervised training. Nonetheless, both approaches produce term-based representations, making LSR compatible with existing infrastructure, such as the inverted indexes originally developed for BM25~\cite{DBLP:journals/corr/abs-1910-10687}.

SPLADE \cite{formal_splade_2021}, a state-of-the-art method, leverages a single pretrained model as an encoder for both query and document by using the masked language modeling (MLM) head of the backbone model. This approach performs term expansion and term weighting in an end-to-end manner. The encoder transforms an input sequence $t$ (either query or document) into a logit matrix $\mathbf{W}$, where $W_{i,j}$ represents the weight of the $i^{\text{th}}$ term in the input sequence for the $j^{\text{th}}$ term in the vocabulary. The output provides a sparse representation for each term in the sequence, assigning a non-negative log-scaled weight to each vocabulary term. This representation is then reduced through column-wise max-pooling, selecting the maximum logit value for each term across the sequence. The term weights are formulated as:
\begin{equation}
  w_{i}(t)= \log \left(1+\max_{j=1}^{L} \operatorname{ReLU}\left(h_{j}^{\top} e_{i}+b_{i}\right)\right)
\end{equation}
where $\operatorname{ReLU}$ ensures non-negativity and the $\log$ normalization prevents excessive weight values.

\subsection{Cased vs. Uncased Models}

The choice between cased and uncased pre-trained models has been widely studied in NLP. BERT \cite{devlin_bert_2019} was originally released in both cased and uncased versions, with the recommendation to use the cased variant for tasks where capitalization carries semantic information (e.g., NER), and the uncased variant for more general applications such as open-domain question answering.

In the early stages of Neural IR, both dense methods (e.g., DPR~\cite{dai_deeper_2019}, ColBERT~\cite{khattab_colbert_2020}) and sparse methods (e.g., SPARTA~\cite{zhao_sparta_2020}, SPLADE~\cite{formal_splade_2021}) predominantly employed uncased versions of models such as BERT and DistilBERT, largely due to the focus on open-domain document retrieval where benchmarks like MS MARCO~\cite{bajaj_ms_2018} were widely evaluated. Despite their prevalence, no explicit comparisons between cased and uncased models were conducted in IR, and to the best of our knowledge, no systematic evaluation has been reported, leaving uncased models as the dominant choice. By contrast, NLP research has investigated this issue more directly; for example, \cite{laskar_utilizing_2020} showed that BERT-cased can achieve sufficient performance in open-ended question answering when pre-processing is applied.

Cased models have also been shown to be beneficial in domain-specific text mining tasks where capitalization encodes critical semantics. This is observed by BioBERT~\cite{lee_biobert_2020}, which leverages cased inputs to achieve superior performance on biomedical corpora. While DyVo~\cite{nguyen_dyvo_2024} extended SPLADE's vocabulary to incorporate case-sensitive named entity terms, these named entity terms were used only as opaque expansion terms, meaning that their casing played no role in whether they were activated and the model's input remained uncased.

Recent evidence also suggests that cased models can improve dense retrieval for open-domain question answering. For example, ModernBERT \cite{warner_smarter_2024} was compared with both base and large variants of uncased BERT (\textit{BERT-base-uncased}, \textit{BERT-large-uncased}), showing performance gains with certain dense retrieval techniques. Dense retrieval architectures allow relatively straightforward switching between cased and uncased models, as similarity is computed over dense vectors rather than token-level matches. In contrast, in LSR, where token-level matching is central, the impact of capitalization is more intricate and remains less studied.

More recently, researchers have begun extending LSR beyond encoder-only architectures, exploring encoder-decoder and decoder-only models \cite{qiao_leveraging_2025,xu_csplade_2025,zeng_scaling_2025,doshi_mistral-splade_2024}. These approaches generally rely on decoder backbones, which, unlike BERT, are typically released only in cased form. While this shift demonstrates that decoders can be effective for sparse retrieval, it also obscures the question of how much capitalization contributes to their performance, since uncased variants are not available for comparison. Consequently, the role of casing in decoder-based LSR remains underexplored. 

In this paper, we focus on encoder-only models that provide both cased and uncased variants. This setting allows us to systematically investigate the effect of capitalization in LSR, including the design of pre-processing and post-processing strategies that might exploit cased inputs. By isolating these effects, we aim to clarify whether capitalization has a consistent effect on LSR; a question that remains open despite its potential significance for both encoder- and decoder-based retrieval.

\section{Methodology}\label{sec:method}
\begin{figure*}[t]
    \centering
    \includegraphics[width=\linewidth]{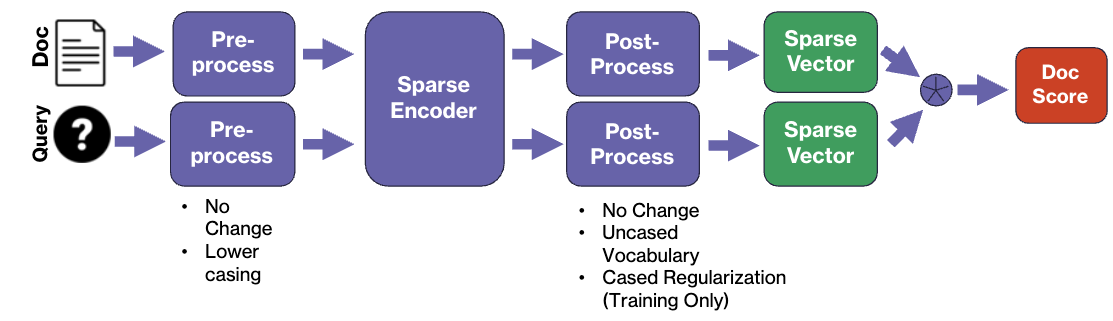}
    \caption{Pipeline of cased models. Queries and documents first undergo a pre-processing step, followed by encoding, and then a post-processing step where sparse vectors are generated and compared. During post-processing, Cased Regularization is applied only during training as an additional loss.}
    \label{fig: cased models pipeline}
\end{figure*}

To evaluate the influence of casing and vocabulary space on model performance, we implemented a set of \textbf{pre-processing} and \textbf{post-processing} strategies, as illustrated in Figure \ref{fig: cased models pipeline}. These modifications were applied at different stages of training and evaluation to control the vocabulary space and avoid term frequency mismatches.
 
\subsubsection{Pre-processing.} We have introduced two techniques that are applied before passing the \textbf{input text} to the model:

\begin{enumerate}
    \item \textbf{None}: The input text is used in its original form. For instance, case words such as ``Apple'' are preserved without modification.

    \item \textbf{Lowercasing}: All input tokens are converted to lowercase. For example, ``Apple'' becomes ``apple''.
\end{enumerate}

\subsubsection{Post-processing.} We have introduced three techniques that are applied to the \textbf{output embeddings} of the model:

\begin{enumerate}
    \item \textbf{None}: The full output embedding is retained, preserving both cased and uncased token dimensions.

    \item \textbf{Uncased Vocabulary Only}: Only the embedding dimensions corresponding to uncased tokens are preserved; those associated with cased tokens are zeroed. Words that exist exclusively in cased form are treated as their lowercase equivalents.

    \item \textbf{Cased Regularizer (Training Only)}: During training, a regularization term is introduced to discourage the model from relying on cased token embeddings. This term penalizes the L2 norm of activations corresponding to cased dimensions. A polynomial warm-up schedule is used to ensure stable optimization. The regularization loss is defined as:

    \begin{equation}
        \mathcal{L}_{\text{Cased}} = w_t \sum_{j=1}^{N} (a_{\mathcal{I}_c}^j)^2  = w_t \sum_{j=1}^{N} \sum_{i \in \mathcal{I}_c} \left( a^j_i \right)^2,
        \label{eq:cased_reg}
    \end{equation}
    where $w_t$ is the time-dependent regularization weight, $N$ is the batch size, $\mathcal{I}_c$ is the set of indexes of the cased token dimensions, $a_i^j$ is the $i$-th activation of the $j$-th sample, and $a_{\mathcal{I}_c}^j$ collects the activations over all cased dimensions for sample $j$.

\end{enumerate}

\section{Experimental Setup}\label{sec:exp-setup}

\subsubsection{Datasets.}
We trained and evaluated our models on the \textbf{MSMARCO Passage V1}~\cite{bajaj_ms_2018} dataset, which contains approximately $8.8 \times 10^6$ passages. Additionally, we assessed model performance on the \textbf{TREC Deep Learning 2019} (DL-2019)\cite{DBLP:journals/corr/abs-2003-07820} and \textbf{TREC Deep Learning 2020} (DL-2020) benchmarks~\cite{DBLP:conf/trec/CraswellMMYC20}, both derived from the MSMARCO Passage V1 corpus and annotated with professionally judged relevance labels. To further test generalization, we evaluated our models on the \textbf{BEIR}~\cite{lee_biobert_2020} benchmark to assess out-of-distribution performance across diverse retrieval tasks.

\subsubsection{Models.}
We evaluated the following models:
\begin{enumerate}
    \item \textbf{BM25} \cite{robertson_simple_1994}: A statistical retrieval method that operates on a bag-of-words representation and requires no training or encoder.
    \item \textbf{SPLADE-BERT} \cite{devlin_bert_2019}: A SPLADE model trained using BERT as the backbone. BERT is a deep bidirectional transformer encoder pre-trained with masked language modeling (MLM) and next sentence prediction (NSP). We fine-tuned publicly available BERT-base uncased\footnote{\url{https://huggingface.co/google-bert/bert-base-uncased}} and cased\footnote{\url{https://huggingface.co/google-bert/bert-base-cased}} checkpoints.
    \item \textbf{SPLADE-DistilBERT} \cite{sanh_distilbert_2020}: A SPLADE model trained using DistilBERT as the backbone. DistilBERT is a lightweight variant of BERT, obtained through knowledge distillation, that is 40\% smaller and 60\% faster while retaining 97\% of the original's performance. We used both the original uncased\footnote{\url{https://huggingface.co/DistilBERT/DistilBERT-base-uncased}} and cased\footnote{\url{https://huggingface.co/DistilBERT/DistilBERT-base-cased}} versions.
\end{enumerate}

\subsubsection{Training.}
All models were trained using a \textbf{teacher-student distillation framework} with the \textbf{Margin-MSE loss}\footnote{Teacher scores used to compute the pairwise margins are provided with \url{https://www.dropbox.com/s/sl07yvse3rlowxg/vienna_triplets.tar.gz?dl=0}.}. In this approach, each query is paired with a positive and a negative document, and the student model learns to match the \emph{relative margin} provided by the teacher model. Formally, for a query $q$ and a positive ($d^+$) and negative ($d^-$) document pair, the Margin-MSE loss is defined as:

\begin{equation}
\mathcal{L}_{\text{Margin-MSE}} = \frac{1}{B} \sum_{i=1}^{B} 
\bigl[ f_S(q_i,d_i^+) - f_S(q_i,d_i^-) - f_T(q_i,d_i^+) + f_T(q_i,d_i^-) \bigr]^2
\label{eq:MarginMSE}
\end{equation}
where $B$ is the batch size, $f_S$ and $f_T$ denote the student and teacher model scores respectively. This formulation encourages the student model to preserve the teacher's ranking of positive and negative documents.

To improve representation efficiency, we applied \textbf{FLOPs-based regularization}~\cite{paria_minimizing_2020} to promote sparsity in the query and document embeddings as introduced in Equation \eqref{eq: FLOPs}:

\begin{equation}
  \text{FLOPs} = \sum_{i=1}^{|V|} a_{i}^{2} = \sum_{i=1}^{|V|} \left( \frac{1}{N} \sum_{j=1}^{N} a_{i}^{j} \right)^{2}
  \label{eq: FLOPs}
\end{equation}
where \({a}_{i} \) represents the estimated activation probability of the \( i \)-the dimension across the batch.

\subsubsection{Retrieval Evaluation.}
We assessed retrieval effectiveness using MRR@10 and R@1000 for the MSMARCO dataset, and nDCG@10 and R@1000 for all other datasets. Additionally, we measured FLOP loss on a subset of MSMARCO DEV queries to estimate efficiency, following the methodology used in SPLADE~\cite{formal_splade_2021}.

\subsubsection{Implementation Details}
Our implementation is based on the publicly available SPLADE repository\footnote{\url{https://github.com/naver/splade/tree/main}}. All hyperparameters follow those used in SPLADE-v2~\cite{formal_splade-v2_2021}. For the introduced cased regularization term, we set the $\lambda$ value to 0.2 for both queries and documents, using a scheduler to gradually increase it until 10,000 steps were reached, where it remained stable.

\section{Results}\label{sec:results}

\begin{table}[tb!]
  \caption{Results on the MS MARCO Passage Collection (Dev, DL-2019, DL-2020). Models differ in pre-processing, post-processing, and query casing. Evaluation metrics are MRR@10, nDCG@10, and R@1000. Uncased models are trained on the original dataset with lowercasing only. Bold values indicate the best results.}
  \centering
  \resizebox{\textwidth}{!}{
  \small
  \begin{tabular}{c cc cc cc cc c}
  \toprule
  \multirow{2}{*}{\textbf{Model}} & \multirow{2}{*}{\makecell[c]{\textbf{pre-}\\\textbf{processing}}} & \multicolumn{1}{c}{\multirow{2}{*}{\makecell[c]{\textbf{post-}\\\textbf{processing}}}}
  & \multicolumn{2}{c}{\textbf{MSMarco Dev
  }} & \multicolumn{2}{c}{\textbf{DL-2019}} & \multicolumn{2}{c}{\textbf{DL-2020}} & \multirow{2}{*}{\textbf{FLOPS}} \\
  & & \multicolumn{1}{c}{ }& MRR@10 & \multicolumn{1}{c}{R@1000} & nDCG@10 & \multicolumn{1}{c}{R@1000} & nDCG@10 & \multicolumn{1}{c}{R@1000} & \\
  \midrule
  BM25\tiny -RM3 & \makecell[c]{Lowering} & \makecell[c]{Uncased\\ Voc. only} & 18.67 & 87.81 & 47.95 & 73.62 & 49.36 & 75.12 & N/A \\ 
  \midrule
  \makecell[c]{SPLADE\\BERT\\uncased}& \makecell[c]{Lowering} & \makecell[c]{Uncased\\ Voc. only} & 35.77 & \textbf{97.83} & \textbf{74.32} & \textbf{86.73} & \textbf{69.75} & 82.08 & 3.16 \\
  \midrule
   \multirow{12}{*}{\makecell[c]{SPLADE\\BERT\\cased}} & \makecell[c]{None} & \makecell[c]{None} & 31.71 & 96.84 & 68.09 & 80.67 & 62.39 & 73.35 & 4.43 \\
  \cmidrule{2-10}
   & \makecell[c]{None} & \makecell[c]{Uncased\\Voc. only} & 30.89 & 96.73 & 66.44 & 80.72 & 61.00 & 73.17 & 3.59 \\
  \cmidrule{2-10}
   & \makecell[c]{None} & \makecell[c]{Cased\\Reg.} & 30.00 & 96.35 & 69.32 & 79.88 & 60.35 & 71.66 & 2.80 \\
  \cmidrule{2-10}
   & \makecell[c]{Lowering} & \makecell[c]{None} & 35.45 & 97.60 & 72.86 & 85.19 & 69.13 & 80.51 & 3.70  \\
  \cmidrule{2-10}
    & \makecell[c]{Lowering} & \makecell[c]{Uncased\\Voc. only} & 35.42 & 97.58 & 71.80 & 84.90 & 69.35 & 80.40 & 3.04\\
  \cmidrule{2-10}
   & \makecell[c]{Lowering} & \makecell[c]{Cased\\Reg.} & 35.27 & 97.49 & 71.82 & 84.74 & 67.19 & 80.78 & \textbf{1.99} \\
  \midrule
  \makecell[c]{SPLADE\\DistilBERT\\uncased} & \makecell[c]{Lowering} & \makecell[c]{Uncased\\ Voc. only} & \textbf{35.91} & 97.82 & 73.29 & 86.04 & 69.58 & \textbf{82.78} & 2.95 \\
  \midrule
  \multirow{12}{*}{\makecell[c]{SPLADE\\DistilBERT\\cased}} & \makecell[c]{None} & \makecell[c]{None} & 31.32 & 96.75 & 66.67 & 68.60 & 62.29 & 64.31 & 4.73 \\
  \cmidrule{2-10}
   & \makecell[c]{None} & \makecell[c]{Uncased\\Voc. only} & 31.06 & 96.78 & 65.42 & 81.09 & 62.78 & 73.68 & 4.10 \\
  \cmidrule{2-10}
  & \makecell[c]{None} & \makecell[c]{Cased\\Reg.} & 29.87 & 84.46 & 66.95 & 79.32 & 62.40 & 72.80 & 2.79 \\
  \cmidrule{2-10}
   & \makecell[c]{Lowering} & \makecell[c]{None} & 35.69 & 97.56 & 69.95 & 85.50 & 68.56 & 81.10 & 4.46 \\
  \cmidrule{2-10}
   & \makecell[c]{Lowering} & \makecell[c]{Uncased\\Voc. only} & 35.46 & 97.45 & 70.04 & 84.74 & 68.58 & 81.25 & 3.30 \\
  \cmidrule{2-10}
   & \makecell[c]{Lowering} & \makecell[c]{Cased\\Reg.} & 35.49 & 97.35 & 70.50 & 84.34 & 67.98 & 67.98 & 2.04 \\
  \bottomrule
  \end{tabular}
  }
  \label{tab:main_results}
\end{table}

We conduct experiments to evaluate the effect of casing on model performance using both in-domain datasets (MSMARCO, TREC DL-2019, TREC DL-2020, Table~\ref{tab:main_results}) and out-of-domain datasets (BEIR, Table~\ref{tab:beir}). Across both BERT and DistilBERT, uncased models consistently outperform cased counterparts on in-domain tasks. This effect likely arises because MSMARCO-v1 queries are uncased and, as an open-domain corpus, MSMARCO-v1 provides limited utility for capitalization in disambiguation.

Pre-processing plays a crucial role in cased model performance. Without lowercasing, cased models exhibit a performance drop of about $5\%$ relative to the uncased baseline ($\sim 36\%$). Applying lowercasing largely mitigates this gap, raising cased model accuracy to approximately $35.5\%$. For DistilBERT, the residual difference between the uncased variant and the cased variant with lowercasing as pre-processing is negligible ($\sim 0.30\%$). However, lowercasing does not improve efficiency, as DistilBERT with cased-lowering is less efficient than the uncased variant (4.46 vs.\ 2.95).

Post-processing strategies, such as constraining logits, reduce FLOPs by nearly $50\%$ but slightly decrease accuracy ($\sim 0.20\%$ MRR@10). While uncased vocabulary selection slightly improves accuracy compared to cased regularization (e.g., DistilBERT without pre-processing: $31.06\%$ vs. $29.87\%$ MRR@10), cased regularization has better efficiency (4.10 vs. 2.79 FLOPs). Both approaches, however, introduce additional complexity that limits their practical adoption.

\begin{table}[tb!]
  \centering
  \scriptsize

    \caption{Results on the BEIR benchmark datasets. Models include BM25, BERT, and DistilBERT with cased and uncased variants. Cased variants are under different pre-processing conditions. The evaluation metric is nDCG@10. Bold values indicate the best results per dataset.}

  \resizebox{\textwidth}{!}{
    
  \begin{tabular}{l c| ccc | ccc}
  \toprule
  \multirow{3}{*}{\textbf{Corpus}} & \multirow{1}{*}{\textbf{BM25}}  
    & \multicolumn{3}{c|}{\textbf{SPLADE-BERT}} 
    & \multicolumn{3}{c}{\textbf{SPLADE-DistilBERT}} \\ 
  \cmidrule(lr){2-8}
    & \textbf{RM3} & \textbf{Uncased} & \makecell[c]{\textbf{Cased}\\ no pre-process} & \makecell[c]{\textbf{Cased}\\ lowering} 
      & \textbf{Uncased} & \makecell[c]{\textbf{Cased}\\ no pre-process} & \makecell[c]{\textbf{Cased}\\ lowering} \\ 
  \midrule 
  MS MARCO & 47.95 & \textbf{74.32} & 68.09 & 72.86  & 73.29  & 66.67 & 69.95 \\
  ArguAna & 34.24 & 50.44 & 47.96 & 49.72 & \textbf{52.37} & 46.59 & 50.81 \\
  Climate-FEVER & 12.81 & \textbf{23.18} & 21.26 & 22.12 & 22.36 & 21.40 & 21.62 \\
  DBPedia & 27.44 & \textbf{42.95} & 34.79 & 40.81 & 42.23 & 34.39 & 40.80 \\
  FEVER & 42.73 & \textbf{79.95} & 75.95 & 77.37 & 77.98 & 76.70 & 76.87 \\
  FiQA-2018 & 25.26 & \textbf{33.41} & 30.47 & 32.66 & 32.06 & 30.04 & 32.33 \\ 
  HotpotQA & 51.28 & \textbf{68.74} & 62.32 & 66.03 & 68.42 & 63.17 & 65.83 \\ 
  NFCorpus & 32.22 & 32.68 & 30.53 & \textbf{33.32} & 32.97 & 30.07 & 32.66 \\ 
  NQ & 23.09 & \textbf{52.49} & 48.24 & 51.12 & 51.48 & 47.77 & 50.54 \\ 
  Quora & 76.76 & 82.50 & 81.60 & \textbf{82.72} & 81.91 & 80.75 & 81.65 \\
  SCIDOCS & 14.71 & 15.46 & 12.34 & 14.85 & \textbf{15.48} & 11.96 & 15.23 \\
  SciFact & 67.22 & \textbf{69.16} & 65.56 & 68.38 & 67.96 & 65.17 & 67.05 \\ 
  TREC-COVID & 57.61 & \textbf{71.38} & 65.31 & 66.24 & 67.93 & 63.40 & 65.09 \\
  Touché-2020 (v1) & 60.25 & 21.65 & 21.72 & 21.06 & \textbf{67.93} & 19.38 & 20.04 \\ 
  \midrule
  \textbf{Avg. all} & 40.97 & 51.31 & 47.58 & 48.95 & \textbf{53.88} & 46.96 & 49.32 \\ 
  \textbf{Avg. zero-shot} & 40.43 & 49.54 & 46.00 & 48.18 & \textbf{52.39} & 45.45 & 47.73 \\
  \textbf{Best on dataset} & 0 & \textbf{9} & 0 & 2 & 3 & 0 & 0 \\
  \bottomrule
  \end{tabular}
  }
  \label{tab:beir}
\end{table}

For zero-shot transfer, we evaluate models on the BEIR benchmark (Table~\ref{tab:beir}). Uncased models generally achieve stronger results (BERT-uncased leads on 9 of 14 datasets, and DistilBERT-uncased leads on 3). Nonetheless, cased models with lowercasing remain competitive, surpassing uncased models on NFCorpus (33.32\% vs.\ 32.68\%) and Quora (82.72\% vs.\ 82.50\%). Cased models without pre-processing generally underperform but still surpass BM25 across all datasets, with small gaps ($<5\%$) relative to uncased models in Climate-FEVER, FEVER, and SciFact. Interestingly, BERT-cased without pre-processing outperforms both uncased and cased-lowering variants on Touché-2020, indicating that pre-processing choices influence retrieval effectiveness in specialized domains. Overall, uncased models remain the most reliable choice, though cased models with lowercasing prove effective in certain cases.

\subsection{Answering the Research Questions}
Having established the empirical results, we now return to our \hyperref[RQs]{RQs} and analyze them in light of the findings from both in-domain (MSMARCO, DL-2019, DL-2020) and out-of-domain (BEIR) evaluations. 

\subsubsection{\hyperref[RQ1]{RQ1: Casing Hinders In-Domain Retrieval Performance}} 
Casing consistently hinders performance in in-domain retrieval. On MSMARCO Dev, cased BERT models without pre-processing underperform their uncased counterparts by approximately 4--5\% MRR@10, with similar degradations observed in DL-2019 and DL-2020 with drops of 4--6\% nDCG@10. This suggests that capitalization introduces spurious distinctions that are not beneficial in MSMARCO-style queries, which are generally lowercased. Importantly, this degradation persists across both BERT and DistilBERT, indicating that the effect is robust across encoder scales. 

\subsubsection{\hyperref[RQ2]{RQ2: Lowercasing Mitigates the Negative Impact of Casing}} 
Lowercasing pre-processing substantially mitigates the negative impact of casing. For BERT-cased, lowercasing improves MRR@10 from 31.7 to 35.4 on MSMARCO Dev, reducing the gap to the uncased model to less than 0.3 points. A similar effect is observed for DistilBERT, where the difference between uncased and cased-lowercased models shrinks to $\sim$0.3 MRR@10. These results demonstrate that \textbf{cased models can be made competitive for LSR when paired with lowercasing}. Out-of-domain, cased-lowercased models occasionally surpass uncased ones (e.g., NFCorpus: 33.3 vs.\ 32.7; Quora: 82.7 vs.\ 82.5 nDCG@10), highlighting that lowercasing not only closes the gap but can sometimes exploit residual benefits of cased pretraining.

\subsubsection{\hyperref[RQ3]{RQ3: Post-Processing Improves Efficiency, Not Accuracy}} 
Post-processing techniques, such as restricting logits to uncased vocabularies or applying cased regularization, primarily affect efficiency rather than accuracy. For instance, on MSMARCO Dev, restricting logits reduces FLOPs by nearly 50\% (4.1 $\rightarrow$ 2.8) with a marginal MRR@10 decrease ($<$0.2 points). However, improvements in efficiency come at the cost of added implementation complexity and vocabulary-specific design choices, limiting their practical adoption in new domains or pretrained variants without knowledge of the pre-trained vocabulary (e.g., in BioBERT~\cite{lee_biobert_2020}). Thus, while \textbf{post-processing is a viable efficiency optimization}, it is not an performance-enhancing strategy. 

\subsubsection{\hyperref[RQ4]{RQ4: Uncased Models are More Robust for Zero-Shot Transfer}} 
Uncased models are consistently more robust for zero-shot transfer. On BEIR, BERT-uncased achieves the highest nDCG@10 on 9 of 14 datasets, with a +3.3 point average margin over cased-no-pre-processing (49.5 vs.\ 46.0). DistilBERT-uncased similarly leads on 3 datasets, outperforming cased-no-pre-processing by +6.4 points on average. Though in selected datasets, Cased-lowercased models narrow this gap (e.g., BERT: 49.5 vs.\ 48.2), and they even outperform uncased models (e.g., NFCorpus, Quora). Nonetheless, cased-no-pre-processing models consistently underperform, barely exceeding BM25 on some datasets (e.g., Climate-FEVER, SciFact). Overall, \textbf{uncased models remain the most reliable for zero-shot transfer}, while cased-lowercased models provide a competitive alternative in certain domains, but never a consistently superior one. 

\subsection{Analysis of Cased and Uncased Tokens}

\begin{figure}[t]
    \centering
    \includegraphics[width=\linewidth]{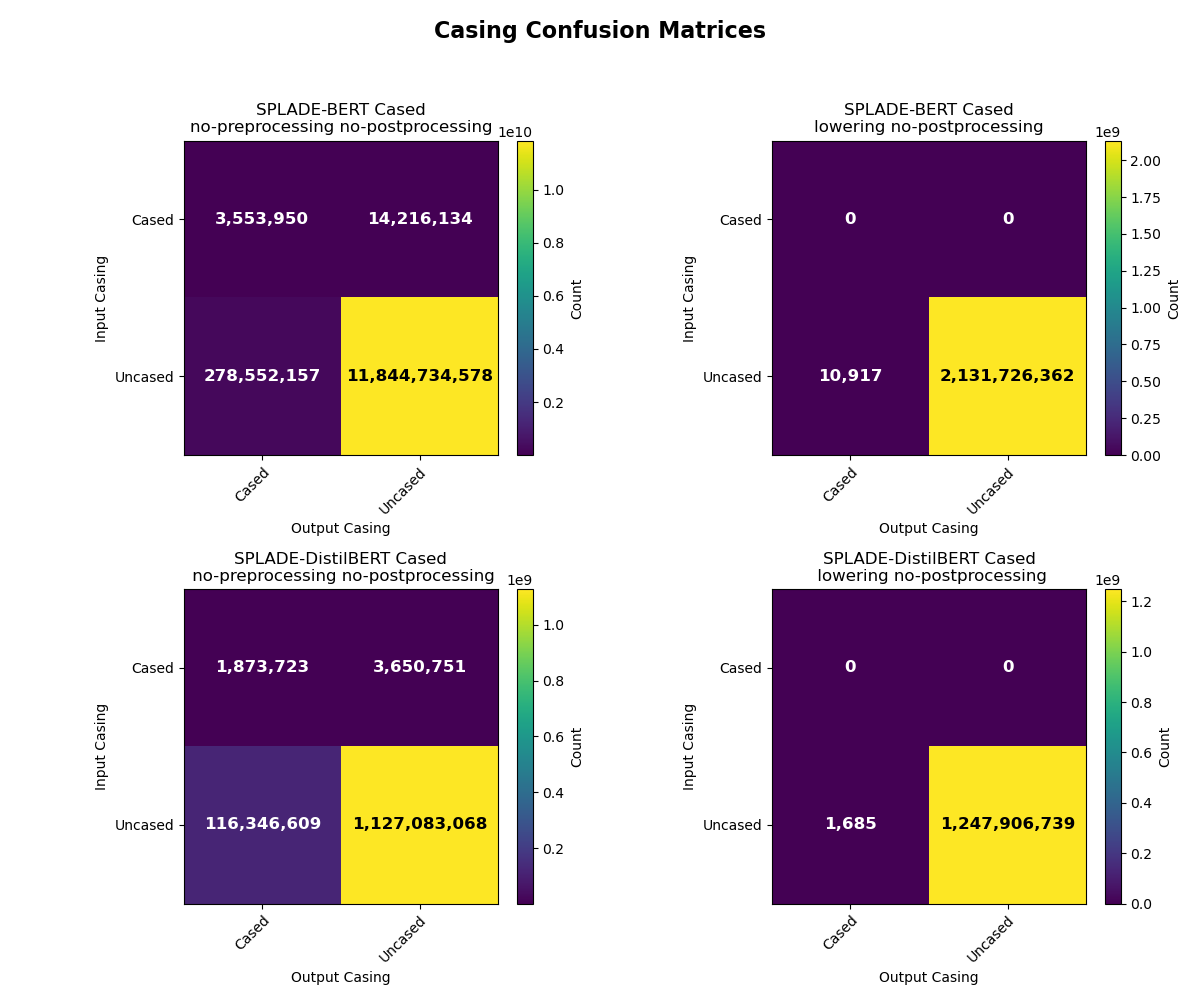}
    \caption{Confusion matrices comparing input and output token casing across BERT and DistilBERT models under different pre-processing conditions. For both models, no post-processing method is used. Rows correspond to input token casing (cased vs. uncased), and columns represent the resulting output token casing. Values denote absolute token counts rather than normalized proportions, highlighting the relative dominance of uncased tokens across conditions.}
    \label{fig:confusion_matrix}
\end{figure}

To assess the role of casing in token generation, we analyzed query representations from the MS MARCO development set, where the majority of queries are lowercased. The goal of this analysis is to quantify the frequency with which input tokens are mapped to their cased or uncased forms in the output. Figure~\ref{fig:confusion_matrix} presents the confusion matrices for BERT and DistilBERT under two pre-processing regimes: (i) no pre-processing and (ii) explicit lowercasing, with no post-processing being applied. 

Across all models and conditions, uncased tokens account for the vast majority of output tokens, surpassing cased tokens by several orders of magnitude. This outcome is not surprising as lowercased words represent both the input data and the vocabulary of the models. The matrices also reveal systematic differences depending on whether pre-processing is applied. When no pre-processing is performed, cased input tokens are more frequently mapped to an uncased token rather than to a cased one. This indicates that even in the presence of cased input, both BERT and DistilBERT exhibit a preference for uncased vocabulary items, likely due to the higher coverage and expressivity of uncased tokens in their vocabularies. Conversely, uncased inputs are overwhelmingly mapped back to uncased outputs, with only a very small proportion ($\approx 0.2\%$) producing cased outputs.  

When input lowercasing is applied, the behavior changes dramatically. In this regime, both BERT and DistilBERT entirely limit the use of cased tokens to less than 1\%, mapping all inputs to uncased outputs. This suggests that, despite being trained with cased vocabularies, the models effectively mimic the behavior of their uncased counterparts when forced into a lowercased input space. Such behavior likely reduces the risk of vocabulary mismatch, though it raises open questions about how the models internally represent and distribute capacity across cased and uncased subspaces.  

Interestingly, in the case of BERT with lowercasing, only two specific tokens appear to be utilized in practice, namely \texttt{R} and \texttt{\#\#R}. These tokens do not correspond to meaningful English words, but rather seem to serve as anchor points within the cased subspace~\cite{DBLP:journals/corr/abs-2110-11540}. This observation warrants further investigation, as it could provide insights into how subword vocabularies adapt to pre-processing constraints and whether this impacts downstream retrieval effectiveness.

\section{Conclusion}\label{sec:concl}

In this work, we systematically examined how the casing of backbone models affects LSR. Our results reveal that while cased models can underperform when used without pre-processing, applying lowercasing effectively recovers most of their retrieval effectiveness, bringing them on par with uncased models. Post-processing techniques, such as leveraging only uncased logits or regularizing cased tokens, offer notable efficiency improvements but only minor gains in accuracy, suggesting that their utility may be task- or resource-dependent. Importantly, these findings indicate that cased models remain viable for LSR, provided that pre-processing strategies are carefully applied. As future work, we want to extend this analysis to more recent cased architectures, like ModernBERT~\cite{warner_smarter_2024}, which could unlock further improvements in both performance and efficiency.

\section*{Acknowledgments}
This research was supported by the Hybrid Intelligence Center, a 10-year program funded by the Dutch Ministry of Education, Culture and Science through the Netherlands Organisation for Scientific Research, and by project VI.Vidi.223.166 of the NWO Talent Programme, which is (partly) financed by the Dutch Research Council (NWO). Additionally, this work benefited from a travel grant provided by the ELLIS Unit Amsterdam through the ELLIS Honors Program, which supported collaboration between the two universities.

\section*{Disclosure of Interests}
The authors have no competing interests to declare that are relevant to the content of this article.

\bibliographystyle{splncs04}
\bibliography{references.dblp}

\end{document}